\documentclass[fleqn,usenatbib]{mnras}

\usepackage{newtxtext,newtxmath}

\usepackage[T1]{fontenc}
\usepackage{ae,aecompl}

\usepackage{graphicx}	
\usepackage{amsmath}	
\usepackage{amssymb}	
\usepackage{natbib}
\usepackage{float}
\usepackage[normalem]{ulem}
\usepackage{xcolor,ulem}
\usepackage{xspace} 
\usepackage{subfig}

\newcommand{\notea}[1]{{\color{white}#1}}  

\newcommand{\kms}{\,km$\,$s$^{-1}$\xspace}	
\newcommand{\teff}{\ensuremath{T_{\mathrm{eff}}}\xspace}
\newcommand{\logg}{\ensuremath{\log g}\xspace}
\newcommand{\rvgsr}{\ensuremath{v_{\mathrm{gsr}}}\xspace}
\newcommand{\feh}{\rm{[Fe/H]}\xspace}

\title[PIGS I: Kinematics]{The Pristine Inner Galaxy Survey (PIGS) I: Tracing the kinematics of metal-poor stars in the Galactic bulge}

\author[A. Arentsen et al.]{
A. Arentsen,$^{1}$
E. Starkenburg,$^{1}$
N. F. Martin,$^{2,3}$ 		
V. Hill,$^{4}$
R. Ibata,$^{2}$
A. Kunder,$^{5}$
\newauthor
M. Schultheis,$^{4}$
K. A. Venn,$^{6}$
D. B. Zucker,$^{7}$ 
D. Aguado,$^{8}$
R. Carlberg,$^{9}$
\newauthor
J. I. Gonz\'alez Hern\'andez,$^{10,11}$
C. Lardo,$^{12}$
N. Longeard,$^{2}$
K. Malhan,$^{13}$
\newauthor
J. F. Navarro,$^{6}$
R. S\'anchez-Janssen,$^{14}$
F. Sestito,$^{1,2}$
G. Thomas,$^{15}$
K. Youakim,$^{1}$
\newauthor
G. F. Lewis,$^{16}$
J. D. Simpson,$^{17}$
Z. Wan$^{16}$
\\
Affiliations are listed after the references
}

\date{Accepted XXX. Received YYY; in original form ZZZ}

\pubyear{2019}

\begin{document}
\label{firstpage}
\pagerange{\pageref{firstpage}--\pageref{lastpage}}
\maketitle

\begin{abstract}
Our Galaxy is known to contain a central boxy/peanut-shaped bulge, yet the importance of a classical, pressure-supported component within the central part of the Milky Way is still being debated. It should be most visible at low metallicity, a regime that has not yet been studied in detail. 
Using metallicity-sensitive narrow-band photometry, the Pristine Inner Galaxy Survey (PIGS) has collected a large sample of metal-poor ($\feh < -1.0$) stars in the inner Galaxy to address this open question. 
We use PIGS to trace the metal-poor inner Galaxy kinematics as function of metallicity for the first time.
We find that the rotational signal decreases with decreasing \feh, until it becomes negligible for the most metal-poor stars. Additionally, the velocity dispersion increases with decreasing metallicity for $-3.0 < \feh < -0.5$, with a gradient of  $-44 \pm 4$\,km$\,$s$^{-1}$\,dex$^{-1}$. 
These observations may signal a transition between Galactic components of different metallicities and kinematics, a different mapping onto the boxy/peanut-shaped bulge for former disk stars of different metallicities and/or the secular dynamical and gravitational influence of the bar on the pressure-supported component. 
Our results provide strong constraints on models that attempt to explain the properties of the inner Galaxy. 
\end{abstract}




\section{Introduction}

The central few kiloparsecs of the Milky Way are a unique place for Galactic studies as there multiple Galactic components overlap with each other. Most of the mass appears to be in a rotation supported boxy/peanut-shaped bulge that rotates like a solid body, indicative of a bar origin \citep{howard09, ness13b}.  In recent years, a debate has been ongoing about the importance of an additional classical, pressure-supported component in the bulge \citep{zoccali08, babusiaux10, kunder16}. If present, it can only be a small percentage of the mass of the bulge, as constrained by the line-of-sight velocity profiles \citep[][]{shen10, ness13b}. However, these profiles are mainly based on metal-rich stars. At lower metallicity the pressure-supported component may be expected to play a larger role. Additionally, the halo continues down into the central regions of our Galaxy and contributes to the metal-poor inner Galaxy. 

Most bulge studies have been based on samples with stars of fairly high metallicity (\feh $> -0.5$), since such stars are the most abundant. They already found that sub-solar metallicity stars are more spherically distributed than super-solar metallicity stars \citep[e.g.][]{zoccali17}, although they show the same amount of rotation \citep{ness13b}. The few studies focussing on metal-poor stars in the inner Galaxy used RR Lyrae stars, a population with mean $\feh = -1.0$ and age $> 11$ Gyr \citep{dekany13,kunder16}. They find that these stars are more spherically distributed than the metal-rich stars and rotate only slightly (if at all). 
The ARGOS red clump bulge survey contains only 4.4 per cent (522 stars) with \feh $< -1.0$  and 0.7 per cent (84 stars) with \feh $< -1.5$ \citep{ness13a}, not sufficient to study this population in detail.

What is missing from the literature is a comprehensive study of the behaviour of the (very) metal-poor tail of the inner Galaxy. In this Letter we present for the first time the kinematics as a function of metallicity for a large sample of metal-poor stars (mainly $\feh < -1.0$) from the Pristine Inner Galaxy Survey (PIGS, Arentsen et al. in prep.).

\vspace{-0.5cm} 

\section{Data}

Since metal-poor stars in the Galactic bulge are extremely rare, targeted selection is necessary to obtain a large sample of them. PIGS is a sub-survey of the Pristine survey \citep{starkenburg17}, which uses a narrow-band $CaHK$ filter for MegaCam on the 3.6m Canada-France-Hawaii Telescope (CFHT) to photometrically search for the most metal-poor stars. The $CaHK$ photometry is highly sensitive to metallicity for FGK stars. Candidate metal-poor stars selected from this photometry are followed up using the AAOmega+2dF multi-fibre spectrograph on the 3.9m Anglo-Australian Telescope (AAT) in low/intermediate resolution, which has 400 fibres in a two-degree field of view. 

\begin{figure}
\centering
\includegraphics[width=0.7\hsize,trim={1.0cm 0.5cm 1.0cm 0.5cm}]{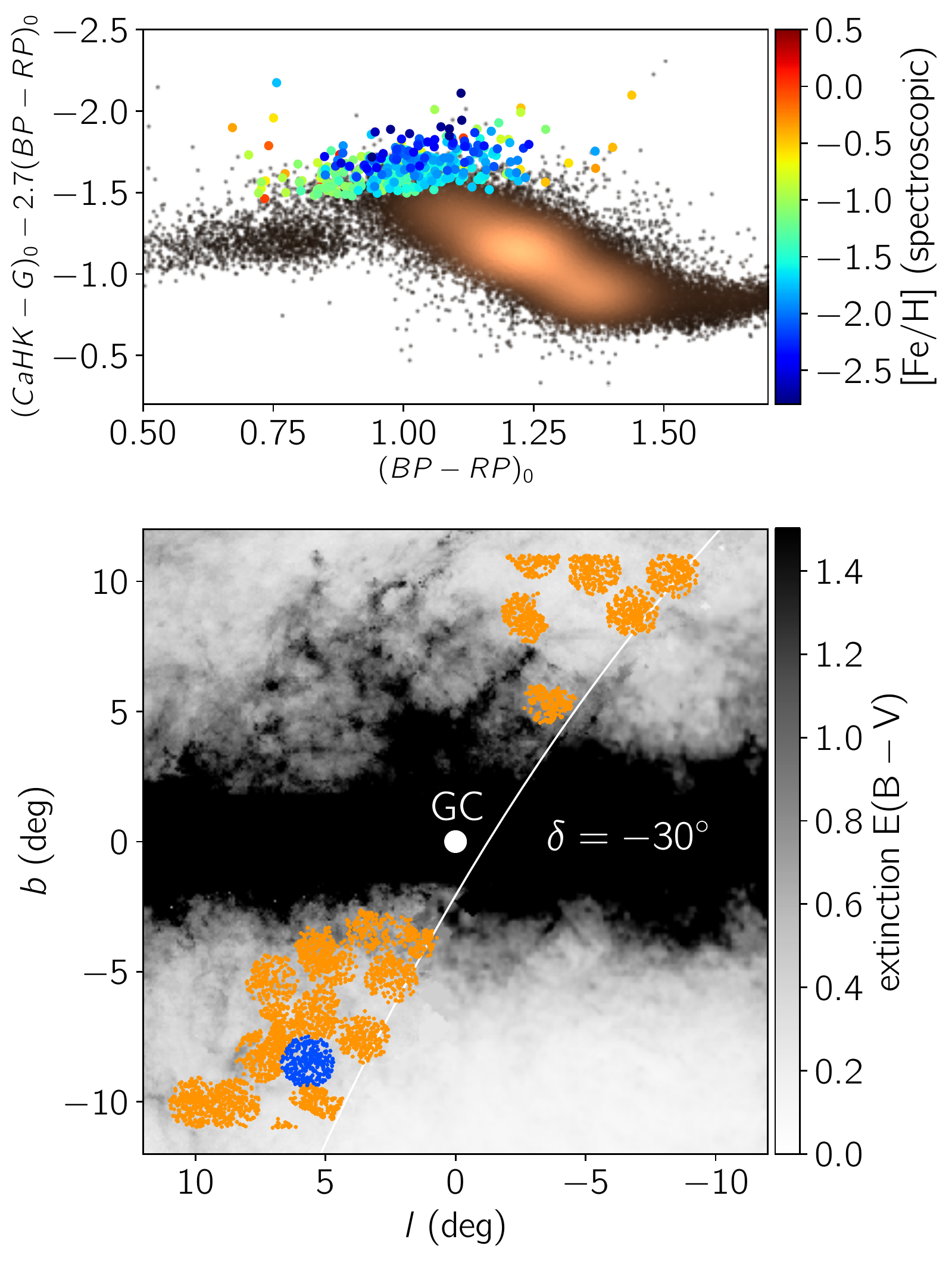}
\caption{
Top: Pristine colour-colour diagram for one field (blue in bottom panel). Metal-poor stars lie towards the top. All stars in this field in our magnitude range that pass the parallax cut are shown ($\sim 40\,000$ stars), of which we observed 350 with the AAT (coloured points).
Bottom: Coverage of the sample of stars used in this work (after quality cuts). The extinction in the background is a combination of the 3D extinction map from \citet{green18} at 8~kpc, available above a declination of $-30^{\circ}$ (white curve) and reliable at $|b| > 2^{\circ}$, and the \citet{schlegel98} extinction map elsewhere. The colour bar is truncated at E(B$-$V) = 1.5 for clarity. 
}
 \label{fig:coverage}
\end{figure}

For the selection, we combine our $CaHK$ photometry with $Gaia$ DR2 broadband $BP$, $RP$ and $G$ photometry \citep{gaia16,gaia18}, or in some later fields with Pan-STARRS1 DR1 $g$ and $r$ photometry \citep{panstarrs}. We checked that there are no strong systematic effects between the two selections that affect the results of this Letter. We deredden the photometry using the 3D extinction map of \citet{green15,green18}, assuming a distance of 8~kpc. To probe giants in the bulge region we select stars with $13.5 < G_0 < 16.5$ ($Gaia$) or $14.0 < g_0 < 17.0$ (Pan-STARRS), excluding those that have $Gaia$ (parallax $-$ parallax$\_$error) > 0.25 mas to avoid stars closer to us than $\sim$4 kpc. We do not make any cuts based on the proper motions. The top panel of Fig.~\ref{fig:coverage} shows the location of spectroscopically followed-up metal-poor candidates in a colour-colour diagram. We select stars moving down from the top of the diagram, until there are enough targets to fill all 2dF fibres.

The spectroscopic observations were performed in August 2017, June 2018, August 2018 and April 2019 with the 580V and 1700D gratings. The blue spectra cover a wavelength range of 3800--5800\,\AA~at a resolution R$\sim$1300, the red spectra cover the calcium triplet (CaT) with a wavelength range of 8350--8850\,\AA~at a resolution R$\sim$10\,000. Total exposure times were generally two hours, with sub-exposures of 30 minutes, to reach a minimum signal to noise ratio (SNR) of 20 for the faintest stars. The data were reduced and combined using the AAT \texttt{2dfdr} pipeline\footnote{https://aat.anu.edu.au/science/software/2dfdr} (v. 6.46) with the standard settings.
The spatial coverage of the sample used in this work is shown in the bottom panel of Fig.~\ref{fig:coverage}. The PIGS footprint only reaches a declination of $-30^{\circ}$ since the $CaHK$ photometry is obtained from the Northern hemisphere. 

We used the \textsc{fxcor} package in IRAF\footnote{IRAF (Image Reduction and Analysis Facility) is distributed by the National Optical Astronomy Observatories, which are operated by the Association of Universities for Research in Astronomy, Inc., under contract with the National Science Foundation.} to determine line-of-sight velocities from the CaT spectra. As templates we use synthetic spectra created using the MARCS~(Model Atmospheres in Radiative and Convective Scheme) stellar atmospheres and the Turbospectrum spectral synthesis code \citep{alvarez98,gustafsson08,plez08}. We first derive velocities using a fixed template, then derive stellar parameters with the zero-shifted spectra, and finally rederive velocities using templates with stellar parameters close to those of each star, in the following grid: $\teff = [5000,5500]$ K, $\logg = 2.5$ and $\feh = [0.0,-1.0,-2.0,-3.0]$. Uncertainties are of order 2~\kms, combining the formal \textsc{fxcor} uncertainties with an estimate of the systematic uncertainties derived from repeated observations. Stars with large line-of-sight velocity uncertainties ($\varepsilon_\mathrm{\textsc{fxcor}} >$ 5 \kms) and double-lined spectra are discarded from our sample. The line-of-sight velocity converted to the Galactic Standard of Rest \rvgsr is determined using \textsc{astropy} \citep[v3.0,][]{astropy13,astropy18}, assuming the peculiar velocity of the Sun from \citet{schonrich10} and circular velocity at the solar radius from \citet{bovy15}.

To determine stellar parameters \teff, \logg and \feh from the blue spectra, we use the full-spectrum fitting University of Lyon Spectroscopic Software (ULySS, \citealt{koleva09}). We use the well-tested model interpolator created from the empirical MILES library \citep{prugniel11,sharma16}, which has for example been used to determine stellar parameters for the X-shooter Spectral Library \citep{arentsen19}. This interpolator is calibrated down to $\feh = -2.8$, which is perfect for the sample in this work. For more details on the stellar parameter determination, see Appendix~\ref{appendix_analysis}. Typical uncertainties on \teff, \logg and \feh 
are 120\,K, 0.35~dex and 0.2~dex, respectively. 

\begin{figure}
\centering
\includegraphics[width=0.8\hsize,trim={0.0cm 0.5cm 0.0cm 0.5cm}]{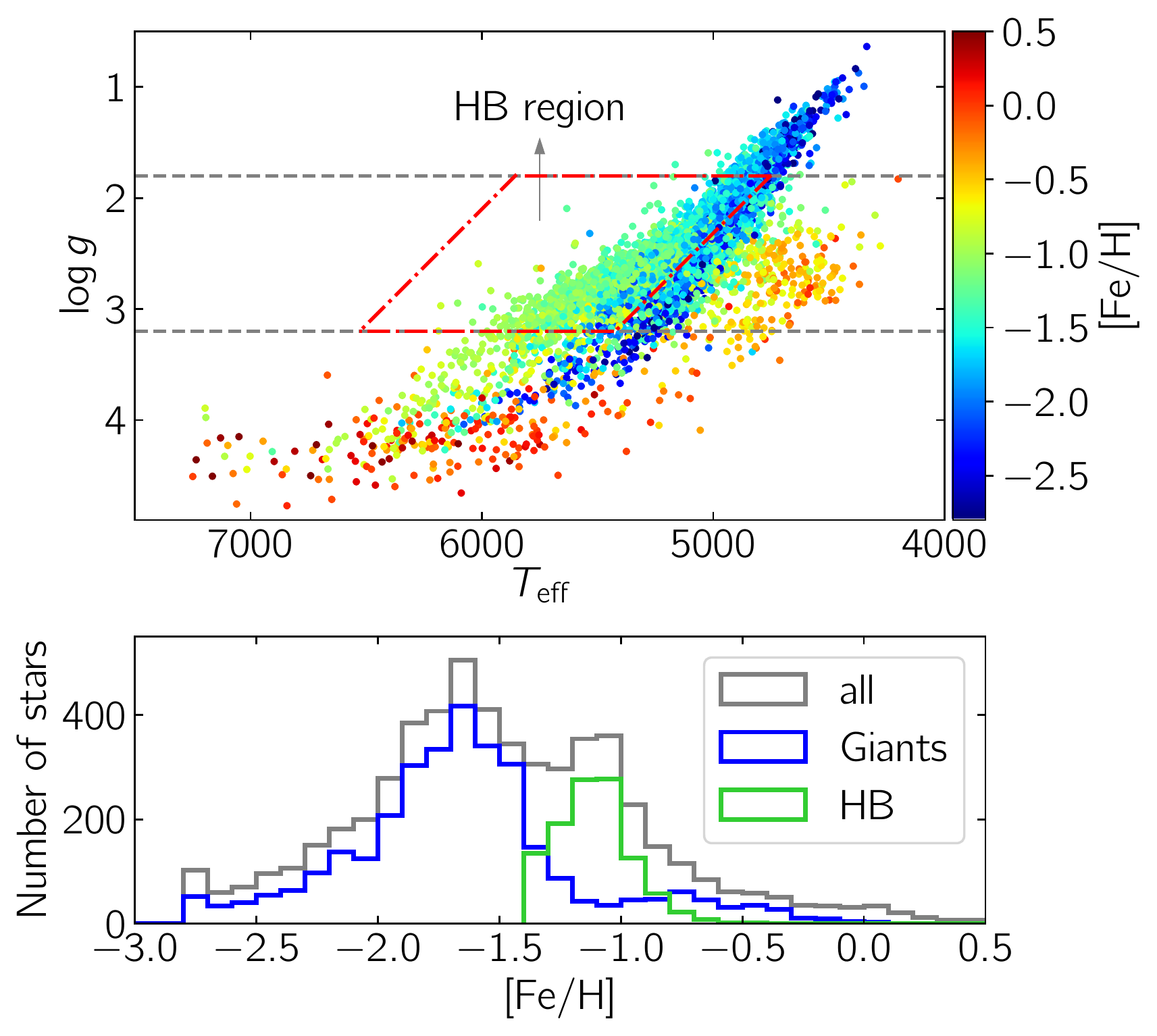}
\caption{ 
Top: \teff--\logg diagram colour-coded by \feh, for the sample making our quality cuts. The the \logg cuts are indicated in grey, and the box containing HB stars is shown in red (see Appendix~\ref{appendix_HB}).
Bottom: Metallicity distribution of our full sample in grey, with the 4200 stars that have $1.8 < \logg < 3.2$ highlighted in blue (giants) and green (HB stars). The lower \feh limit of the interpolator is $-2.8$, causing an apparent excess of stars there.
}
 \label{fig:HRFeH}
\end{figure}

The \teff--\logg diagram of the sample passing our quality criteria is shown in the top panel of Fig.~\ref{fig:HRFeH}. In this Letter, we only consider stars with $1.8 < \logg < 3.2$ (black dotted lines), to constrain the sample to stars likely in the bulge volume. This \logg range is the same as that used in \citet{ness13a} for red clump stars. 
 This cut removes a significant number of the (very) metal-poor stars, since many of those have low surface gravities and may be located further into the halo on the far side of the bulge. For a discussion on the distances of our sample stars, see Appendix~\ref{appendix_dist}. Furthermore, the region where the horizontal branch (HB) stars are located is highlighted in red (see Appendix~\ref{appendix_HB} for more details on the selection and metallicities of the HB stars).

The metallicity distribution of all stars is shown in the bottom panel of Fig.~\ref{fig:HRFeH}. At the low-\feh end there is an excess of stars at the limit of the interpolator; several of these may be even more metal-poor than \feh = $-2.8$.
There are two peaks in the distribution, which can be attributed to two different types of stars: the HB stars and the normal giants. The more metal-rich peak around \feh$\sim -1.1$ is that of the HB stars. Although the exact metallicities for the HB stars are somewhat uncertain, they are clearly distinct from the normal giants. The second peak at \feh~=~$-1.6$ is that of the normal giants in the sample and illustrates the success of PIGS in selecting this rare metal-poor population. The overall shape of the metallicity distribution is the result of our specific photometric selection and does not necessarily represent the metallicity distribution function of the underlying population. 

\begin{figure*}
\centering
\includegraphics[width=\hsize,trim={3.0cm 1.2cm 0.0cm 1.2cm}]{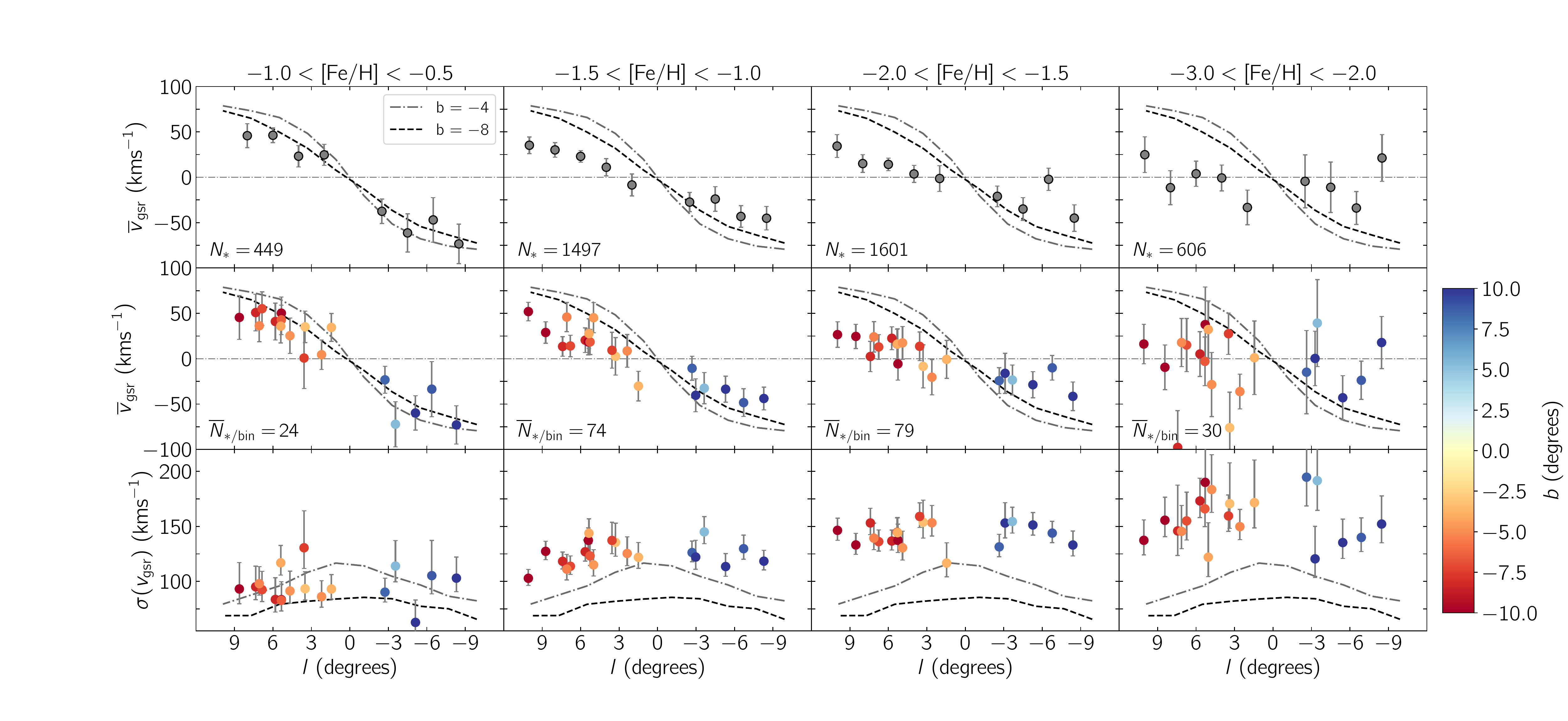}
\caption{Top row: longitude versus mean line-of-sight velocity in the GSR (\rvgsr) for different metallicity ranges from left to right (see column title) in bins of 2 degrees. This includes both the normal giants and the HB stars. Lines from the bar model from \citet{shen10} have been over-plotted. Error bars are $\sigma/\sqrt N$. Middle row: the same, but separately for each AAT field where the colour indicates the latitude. Bottom row: similar, but for the standard deviation of \rvgsr. In each of the panels, only bins with at least 10 stars are shown. The (asymmetric) error bars are $\sqrt{\sigma^2 (N-1)/\chi_{\pm}^2 }$, with $\chi_{\pm}$ determined for a 68\% confidence interval and $N-1$ degrees of freedom.}
 \label{fig:lvsrv}
\end{figure*}

\vspace{-0.3cm} 

\section{Results}

We use our sample to study the kinematics in the inner Galaxy at different metallicities. We investigate the rotation curve and the velocity dispersion as a function of metallicity. 

We present the mean line-of-sight velocity as a function of the Galactic longitude in Fig.~\ref{fig:lvsrv}, with the metallicity decreasing from the left to the right panels. The top row combines all latitudes per longitude bin, the second row shows each AAT field individually, colour-coded by latitude. The dashed lines in these rows are the rotation curves from \citet{shen10}, who modelled a purely barred bulge which fits the BRAVA data \citep{howard08,howard09}. In the bottom row of the figure we present the corresponding velocity dispersions for both the data and the bar model. 

The motions of stars in the most metal-rich bin ($-1.0 < \feh < -0.5$) appear quite consistent with the bar models. The more metal-poor stars also show signatures of rotation, but the mean velocity is smaller and decreases with decreasing metallicity until it disappears completely for the most metal-poor stars ($\feh < -2.0$). 

The rotation of metal-poor stars in the inner Galaxy has previously been seen in the ARGOS data \citep{ness13b}, although their sample of metal-poor stars is small and has only one bin ($\feh < -1.0$). They find line-of-sight velocities of 20--40 \kms at $|l| = 10^{\circ}$, consistent with what we find in our $-1.5 < \feh < -1.0$ bin. Additionally, \citet{kunder16} showed that the RR Lyrae stars in the inner Galaxy may be rotating slowly, although their sample only contains stars with $|l| < 4^{\circ}$ so they cannot trace the rotation beyond this. We clearly see for the first time how the rotation changes, and then disappears, as a function of metallicity in the metal-poor tail of the inner Galaxy.

\begin{figure}
\centering
\includegraphics[width=1\hsize,trim={0.5cm 0.7cm 00cm 0.5cm}]{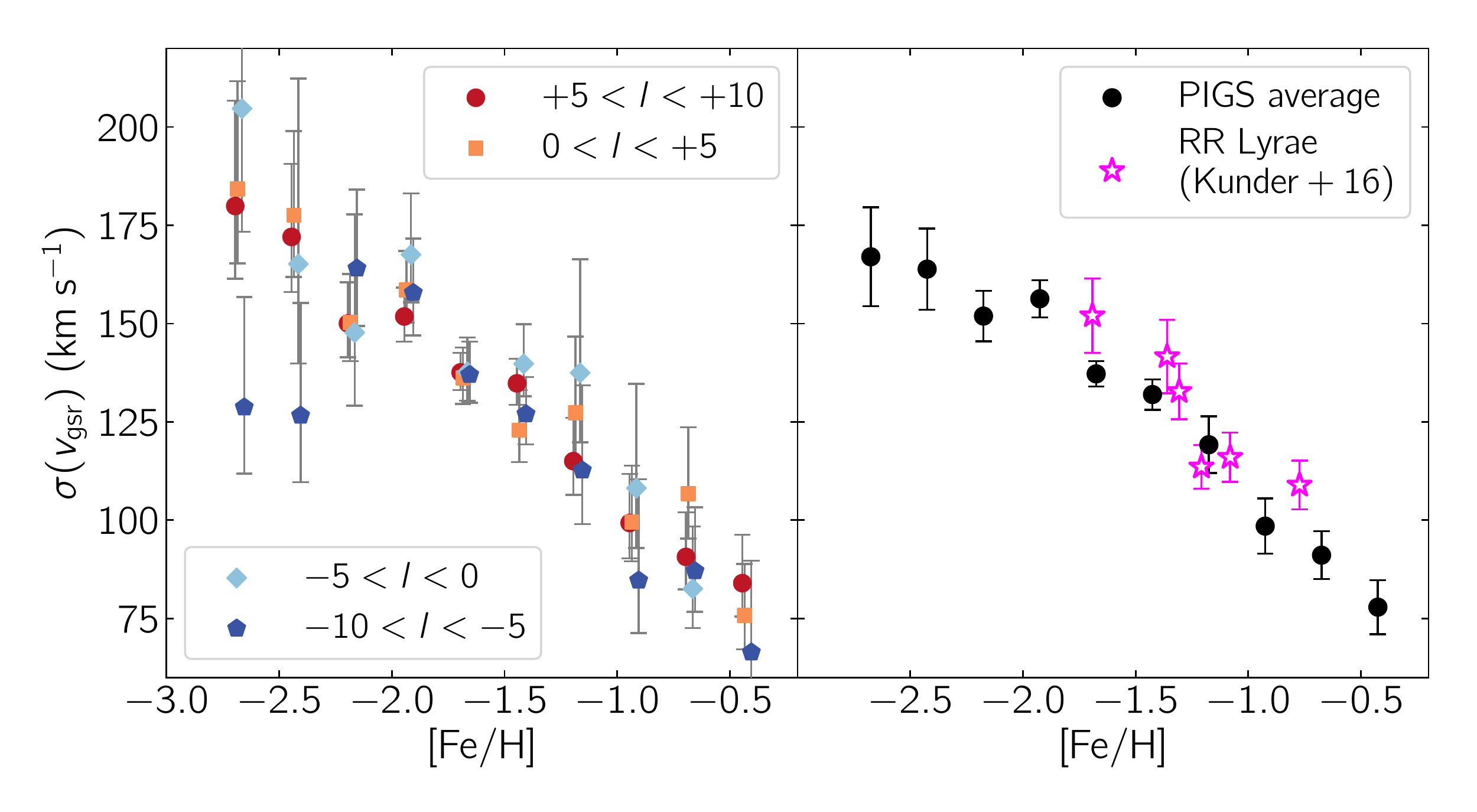}
\caption{
Left: \rvgsr dispersion as a function of \feh (bins of 0.25 dex), for four ranges in $l$. The HB stars are excluded. Different $l$ ranges are offset by 0.01 dex for clarity. Only bins with at least 10 stars are shown. 
Right: weighted average of the different $l$ ranges, with the bulge RR Lyrae results from \citet{kunder16}.
}
 \label{fig:fehvsdisp}
\end{figure}

Additionally, it is striking how strongly the velocity dispersion increases with decreasing metallicity. The behaviour of the velocity dispersion as a function of \feh is shown in more detail in Fig.~\ref{fig:fehvsdisp}. Here we excluded the HB stars, since their metallicity is less well-determined. The increase in velocity dispersion with decreasing \feh appears gradual. The lower dispersions in the range $-10^{\circ} < l < -5^{\circ}$ with $\feh < -2.3$ could be an indication of substructure, but the sample is small. A linear fit to the average velocity dispersions provides a gradient of $-44 \pm 4$\,km$\,$s$^{-1}$\,dex$^{-1}$. 

 \citet{kunder16} have previously shown that for inner Galaxy RR Lyrae stars the velocity dispersion increases with decreasing metallicity. Their results have been over-plotted in the right panel of Fig.~\ref{fig:fehvsdisp} and they generally agree with our findings, although their dispersions appear slightly higher.
 \citet{ness13b} also investigate the velocity dispersion as a function of \feh, with lower \feh resolution and mainly for higher metallicity stars. For fields with $|l| \leqslant 10^{\circ}$ and $b \leqslant -7.5^{\circ}$, their relations appear to have similar slopes between $-1.0 < \feh < +0.5$. Their velocity dispersion at $\feh = -1.0$ of $\sim100$ \kms is similar to our findings. Combining the results of PIGS and ARGOS, we find an almost continuous increase in velocity dispersion from $\feh = +0.5$ down to $\feh = -3.0$.
 
\vspace{-0.5cm} 

\section{Discussion}

With our large PIGS sample of metal-poor inner Galaxy stars, we can trace for the first time their rotation and velocity dispersion as a function of the metallicity for $-3.0 < \feh < -0.5$. We find that the metal-poor stars show a signature of rotation, which decreases in magnitude with decreasing \feh, until it disappears for the most metal-poor stars. The line-of-sight velocity dispersion is continuously increasing with decreasing metallicity. What is the interpretation of the behaviour of these metal-poor inner Galaxy stars? We present a number of possible scenarios. 

Firstly, we note that our sample is not completely free of selection effects. Metal-poor stars are generally brighter and thus we may preferentially select stars that are further away at lower metallicity (and therefore less dominated by ``the bulge''). Although we have tried to mitigate this effect through our selection of stars based on surface gravity, a small effect is potentially still present.

If we take our results at face value, a possible explanation is a smooth transition of how much each of the Galactic components (with fixed velocity dispersions) contributes at a certain metallicity. In this case there is a change from being dominated by a rotation-supported population of a bar or disc at higher metallicities ($\feh > -1.0$, \citealt{ness13b}) to being fully dominated by the pressure-supported component like the halo or classical bulge at the lowest metallicities ($\feh < -2.0$). Additionally, the stars in our sample are not necessarily confined to the bulge. At the lowest metallicities the fraction of stars that have large apocentres, and therefore only pass through the bulge, may increase. A transition between components can naturally explain an increase in velocity dispersion and a decreasing rotational signal with decreasing metallicity. This scenario would also work if the pressure-supported component itself does not rotate at all. 

It has been shown that in the solar neighbourhood the thick disc has a metal-poor tail with $\feh < -1.5$ \citep{kordopatis13}, possibly extending all the way down to $\feh < -4.0$, \citep{sestito19}. If this tail is also present in the inner Galaxy, these stars could contribute to the rotational signal that we see in Fig.~\ref{fig:fehvsdisp}. \citet{kordopatis13} estimate that in the solar neighbourhood, the fractions of stars with $\feh < -1.5$ belonging to the thick disc or the halo are equal (selecting stars with $1 < |Z/\mathrm{kpc}| < 2$ from DR4 of the RAVE survey, \citealt{kordopatis13b}). If we make the simple assumption that this fraction is the same in the inner Galaxy and we combine this with a halo/thick disc mass fraction of 0.10 locally \citep{kordopatis13} and 0.15 in the inner Galaxy (as derived by \citealt{schiavon17} based on the Besan\c{c}on model within a height of 4 kpc and R$_\mathrm{GC}$ = 2 kpc, from \citealt{robin12,robin14}), we estimate that the halo has $\sim1.5$ times more stars in this region than the thick disc for $\feh < -1.5$. Many assumptions are made here, but this argument shows that at these metallicities the densities of the halo and thick disc only differ by a factor of a few. At higher (lower) \feh, the thick disc (halo) contribution will dominate over the other.

Alternatively, stars of all \feh could originate from the disc without an additional halo component, where stars have been mapped into the boxy/peanut bulge in different ways because of their different velocity distributions at the time of the bar formation \citep[e.g.][]{dimatteo16,debattista17}. This fits with a continuous increase in velocity dispersion all the way from $\feh = +0.5$ down to $\feh = -3.0$ (where this is almost like an age-\feh relation). 

Additionally, a present pressure-supported component can itself be rotating. This can be original rotation from the collapse of a slightly rotating cloud for an in-situ classical bulge/halo, or alternatively stars in the spheroidal component could have been spun up and/or caught by the bar. The strong dynamical and gravitational effects of the bar should secularly affect all populations in the inner Galaxy \citep{saha12,saha15}. Recently, \citet{perezvillegas17} used N-body simulations to study the influence of the bar and boxy/peanut bulge on the central part of the halo. They find that due to angular momentum transfer, an initially non-rotating halo starts rotating with line-of-sight velocity signatures of $\sim$15--25 \kms for $|l| > 5^{\circ}$ with a velocity dispersion of $\sim$120 \kms. They also find that a small fraction of the stars ($\sim$12 per cent) are moving on bar-following orbits at the end of their simulation. The exact values of these numbers depend on the model details, but it shows that there can be some rotation in the halo component which is (much) slower than that of the bar. In order for this scenario to explain our metallicity-dependent results, the velocity dispersions for more metal-poor stars had to have been already (much) higher to begin with. 

To further disentangle this complex region of our Galaxy, and e.g. to distinguish between an accreted and in-situ pressure-supported component, additional information is needed. Better distances (e.g. using a combination of parallaxes, spectroscopy and photometry with the StarHorse code, \citealt{santiago16,queiroz18}) are necessary to derive detailed dynamical properties. Additionally, high-resolution spectroscopic follow-up is necessary to get detailed chemistry of stars, which contains important information not present in metallicities and kinematics alone.


\vspace{-0.4cm}

\section*{Acknowledgements}
	We thank the referee for their helpful comments. We thank Chris Wegg, Tobias Buck and Jan Rybizki for helpful discussions and suggestions, and we thank Gail Zasowski for suggesting the PIGS acronym. We thank the organisers of the ESO conference ``The Galactic Bulge at the crossroads'' in Puc\'on, Chile, in December 2018, where the idea of this Letter was conceived.

	We thank the Australian Astronomical Observatory, which have made these observations possible. We acknowledge the traditional owners of the land on which the AAT stands, the Gamilaraay people, and pay our respects to elders past and present. Based on data obtained at Siding Spring Observatory (via programs S/2017B/01, A/2018A/01, OPTICON 2018B/029 and OPTICON 2019A/045, PI: A. Arentsen). 
	Based on observations obtained with MegaPrime/MegaCam, a joint project of CFHT and CEA/DAPNIA, at the Canada-France-Hawaii Telescope (CFHT) which is operated by the National Research Council (NRC) of Canada, the Institut National des Science de l'Univers of the Centre National de la Recherche Scientifique (CNRS) of France, and the University of Hawaii.

	AA, ES and KY gratefully acknowledge funding by the Emmy Noether program from the Deutsche Forschungsgemeinschaft (DFG). 
	NFM, RI, NL, and FS gratefully acknowledge support from the French National Research Agency (ANR) funded project ``Pristine'' (ANR-18-CE31-0017) along with funding from CNRS/INSU through the Programme National Galaxies et Cosmologie and through the CNRS grant PICS07708.
	FS thanks the Initiative dExcellence IdEx from the University of Strasbourg and the Programme Doctoral International PDI for funding his PhD. This work has been published under the framework of the IdEx Unistra and benefits from a funding from the state managed by the French National Research Agency as part of the investments for the future program. 
	CL thanks the Swiss National Science Foundation for supporting this research through the Ambizione grant number PZ00P2 168065.
	DBZ and JDS acknowledge the support of the Australian Research Council through Discovery Project grant DP180101791.
	JIGH acknowledges financial support from the Spanish Ministry of Science, Innovation and Universities (MICIU) under the 2013 Ram\'on y Cajal program MICIU RYC-2013-14875, and also from the Spanish ministry project MICIU AYA2017-86389-P.	
	Horizon 2020: This project has received funding from the European Union's Horizon 2020 research and innovation programme under grant agreement No 730890. This material reflects only the authors views and the Commission is not liable for any use that may be made of the information contained therein.

	The authors thank the International Space Science Institute, Bern, Switzerland for providing financial support and meeting facilities to the international team ``Pristine''.

	This work has made use of data from the European Space Agency (ESA) mission {\it Gaia} (\url{https://www.cosmos.esa.int/gaia}), processed by the {\it Gaia} Data Processing and Analysis Consortium (DPAC, \url{https://www.cosmos.esa.int/web/gaia/dpac/consortium}). Funding for the DPAC has been provided by national institutions, in particular the institutions participating in the {\it Gaia} Multilateral Agreement. 
	
	The Pan-STARRS1 Surveys (PS1) and the PS1 public science archive have been made possible through contributions by the Institute for Astronomy, the University of Hawaii, the Pan-STARRS Project Office, the Max-Planck Society and its participating institutes, the Max Planck Institute for Astronomy, Heidelberg and the Max Planck Institute for Extraterrestrial Physics, Garching, The Johns Hopkins University, Durham University, the University of Edinburgh, the Queen's University Belfast, the Harvard-Smithsonian Center for Astrophysics, the Las Cumbres Observatory Global Telescope Network Incorporated, the National Central University of Taiwan, the Space Telescope Science Institute, the National Aeronautics and Space Administration under Grant No. NNX08AR22G issued through the Planetary Science Division of the NASA Science Mission Directorate, the National Science Foundation Grant No. AST-1238877, the University of Maryland, Eotvos Lorand University (ELTE), the Los Alamos National Laboratory, and the Gordon and Betty Moore Foundation.


\newpage

\bibliographystyle{mnras}
\bibliography{literature.bib}   


\newpage
	
\section*{Affiliations}
\begin{small}
\textit{
\noindent $^{1}$Leibniz-Institut f\"ur Astrophysik Potsdam (AIP), An der Sternwarte 16, D-14482 Potsdam, Germany\\
$^{2}$Universit\'e de Strasbourg, CNRS, Observatoire astronomique de Strasbourg, UMR 7550, F-67000 Strasbourg, France\\
$^{3}$Max-Planck-Institut f\"ur Astronomie, K\"onigstuhl 17, D-69117 Heidelberg, Germany\\
$^{4}$Universit\'e C\^ote d'Azur, Observatoire de la C\^ote d'Azur, CNRS, Laboratoire Lagrange, Blvd de l'Observatoire, F-06304 Nice, France\\
$^{5}$Saint Martin's University, 5000 Abbey Way SE, Lacey, WA 98503, USA \\
$^{6}$Department of Physics \& Astronomy, University of Victoria, Victoria, BC, V8W 3P2, Canada \\
$^{7}$Department of Physics and Astronomy, Macquarie University, Sydney, NSW 2109, Australia \\
$^{8}$Institute of Astronomy, University of Cambridge, Madingley Road, Cambridge CB3 0HA, UK \\
$^{9}$Department of Astronomy \& Astrophysics, University of Toronto, Toronto, ON M5S 3H4, Canada \\
$^{10}$Instituto de Astrof\'isica de Canarias, V\'ia L\'actea, 38205 La Laguna, Tenerife, Spain \\
$^{11}$Universidad de La Laguna, Departamento de Astrof\'isica, 38206 La Laguna, Tenerife, Spain \\
$^{12}$Laboratoire d'Astrophysique, \'Ecole Polytechnique F\'ed\'erale de Lausanne, Observatoire de Sauverny, 1290 Versoix, Switzerland \\
$^{13}$Oskar Klein Centre for Cosmoparticle Physics, Department of Physics, Stockholm University, AlbaNova, 10691 Stockholm, Sweden \\
$^{14}$UK Astronomy Technology Centre, Royal Observatory Edinburgh, Blackford Hill, Edinburgh, EH9 3HJ, UK \\
$^{15}$NRC Herzberg Astronomy and Astrophysics, 5071 West Saanich Road, Victoria, BC V9E 2E7, Canada \\
$^{16}$Sydney Institute for Astronomy, School of Physics, A28, The University of Sydney, NSW 2006, Australia \\
$^{17}$School of Physics, UNSW, Sydney, NSW 2052, Australia \\
}
\end{small}


\newpage

\appendix

\section{Analysis of the spectra}\label{appendix_analysis}

To determine stellar atmospheric parameters, we use the ULySS code \citep{koleva09} with the empirical MILES spectral library interpolator \citep{prugniel11, sharma16}. One advantage of using an empirical reference grid of stellar spectra in the determination of stellar atmospheric parameters, is that the grid only contains spectra with existing combinations of \teff, \logg and \feh, and therefore one only obtains physically sensible solutions. This especially an advantage at low resolution, where the sensitivity to \logg is small. An extensive description of our analysis will be published in the PIGS II paper (Arentsen et al. in prep.), but some relevant details are summarised here. 

We use only the blue arm in the fit, because the maximum wavelength in the MILES library is 7400~\AA. In the future, we will explore the possibility of fitting together the blue and the red arm with a different reference library. The spectra are corrected for their radial velocities, but they are not continuum-normalised, before feeding them into ULySS. Inside ULySS, the continuum, stellar parameters, line broadening, and residual velocities are fitted simultaneously. In the fit, we mask the CH-band around 4300~\AA, since [C/Fe] varies widely at low \feh~\citep[e.g.][and references therein]{beerschristlieb05}. Additionally, ULySS iteratively clips regions from the fit that are large outliers, excluding e.g. dead pixels and other unphysical features in the spectra. 

We select stars with good fits on the basis of a set of (empirically determined) quality criteria. These are the following: 

\begin{itemize}
\item SNR$_{4000-4100}$ > 7, to keep only spectra that have large enough signal-to-noise-ratio (SNR) between 4000--4100~\AA,
\item signal-to-residual ratio (SRR) > 15, to exclude peculiar stars that have bad fits even though they have a good SNR,
\item $\sigma$ < 150 \kms, to exclude stars that were fit with a high line broadening in ULySS, which usually indicates a bad fit,
\item fraction of pixels clipped-out in the fit between $4600-4800$~\AA~$< 10 \%$ and between $5000-5200$~\AA~also $< 10\%$, this cuts out stars with very large carbon features in these wavelength regions. Carbon is not included as a variable in the model spectra which therefore results in bad fits for very carbon-rich stars,
\item \teff > 4200~K, because we found that the fits below this temperature are badly fitted.
\end{itemize}

These quality cuts remove 10\% of the spectra. External tests of the stellar atmospheric parameters show that the systematic uncertainties with our method at the given resolution are of the order of 120\,K, 0.35~dex and 0.2~dex for \teff, \logg and \feh, respectively. Tests with repeated observations show that the statistical uncertainties for stars with SNR$_{4000-4100}$ < 7 are larger than this, therefore we take this as our lower limit in SNR. Details of the tests will be described in the PIGS II paper.  The mean SNR$_{4000-4100}$ of the full sample is 25. For the spectra passing the defined quality criteria, the SNR in the CaT is larger than 10, with a mean of 60.

\begin{figure*}
\centering
\includegraphics[width=1.0\hsize,trim={0.5cm 0.5cm 0.0cm 0.0cm}]{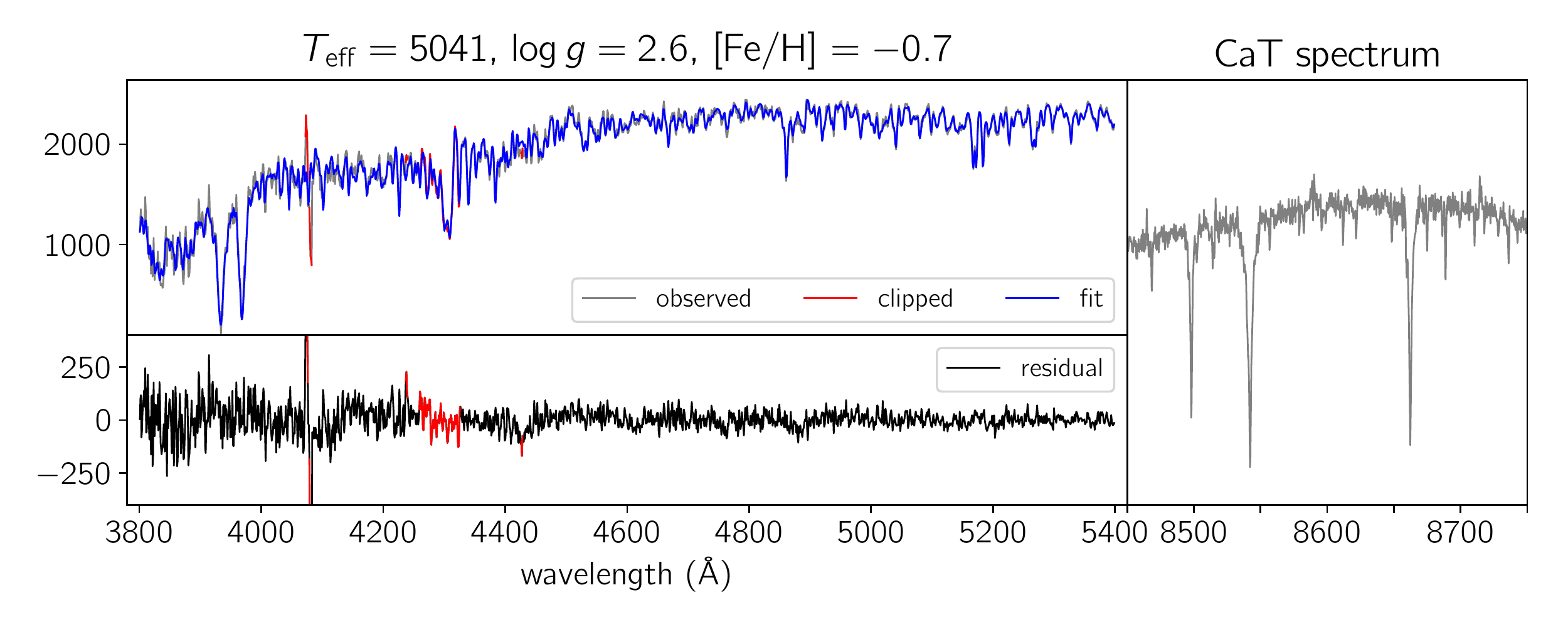}
\includegraphics[width=1.0\hsize,trim={0.5cm 0.5cm 0.0cm 0.0cm}]{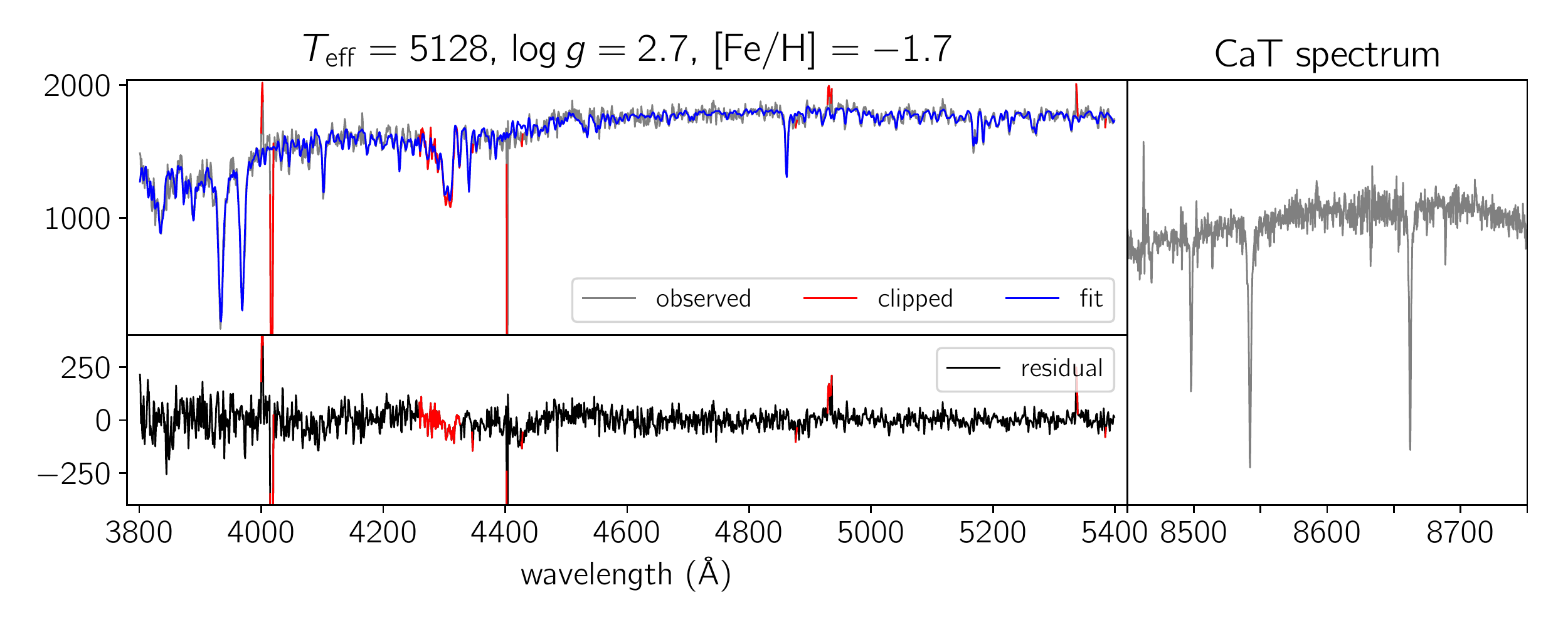}
\includegraphics[width=1.0\hsize,trim={0.5cm 0.5cm 0.0cm 0.0cm}]{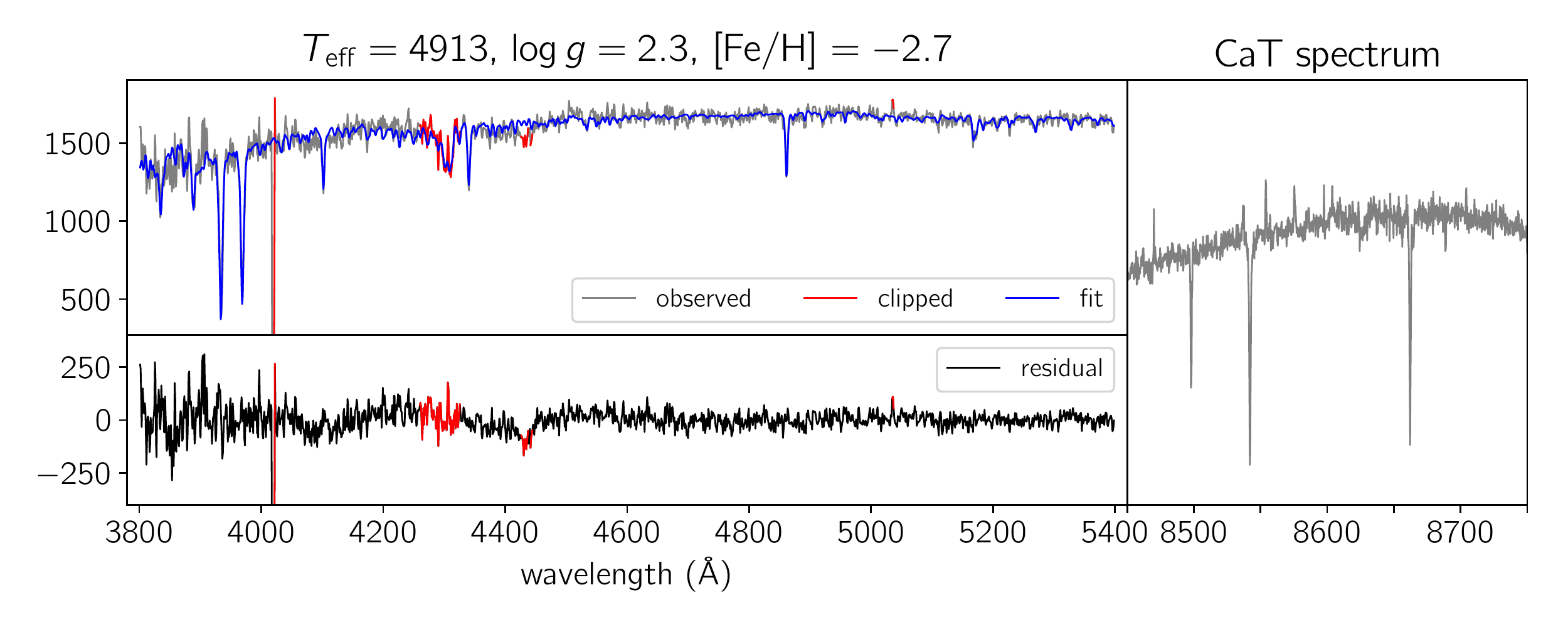}
\caption{Example fits of three spectra at different \feh and slightly different \logg and \teff (see titles). The units on the y-axis are an arbitrary flux unit. The spectra are normalised as determined in the fit. The CH-band around 4300~\AA~has been masked in all fits. The (non-normalised) CaT spectra are shown in the right-hand panels as examples of what they look like, they were not included in the fit.}
 \label{fig:spec}
\end{figure*}

Example fits for three stars of different metallicities are shown in Fig.~\ref{fig:spec}. Each of these spectra has SNR$_{4000-4100}~\approx~25$.

\newpage

\section{Horizontal Branch stars}\label{appendix_HB}

A striking feature in the \teff--\logg diagram (top panel of Fig.~2 in the main paper) is that of the horizontal branch (HB) stars. HB stars are helium-burning stars like red clump stars, but at lower metallicity. They are located at hotter temperatures compared to normal giants of the same metallicity. They enter our selection box on the blue side of the red giant branch (see the top panel of Fig.~1 in the main paper between $0.7 < (BP - RP)_0 < 1.0$). 

\begin{figure*}
\centering
\includegraphics[width=0.9\hsize,trim={0.5cm 0.5cm 0.0cm 0.0cm}]{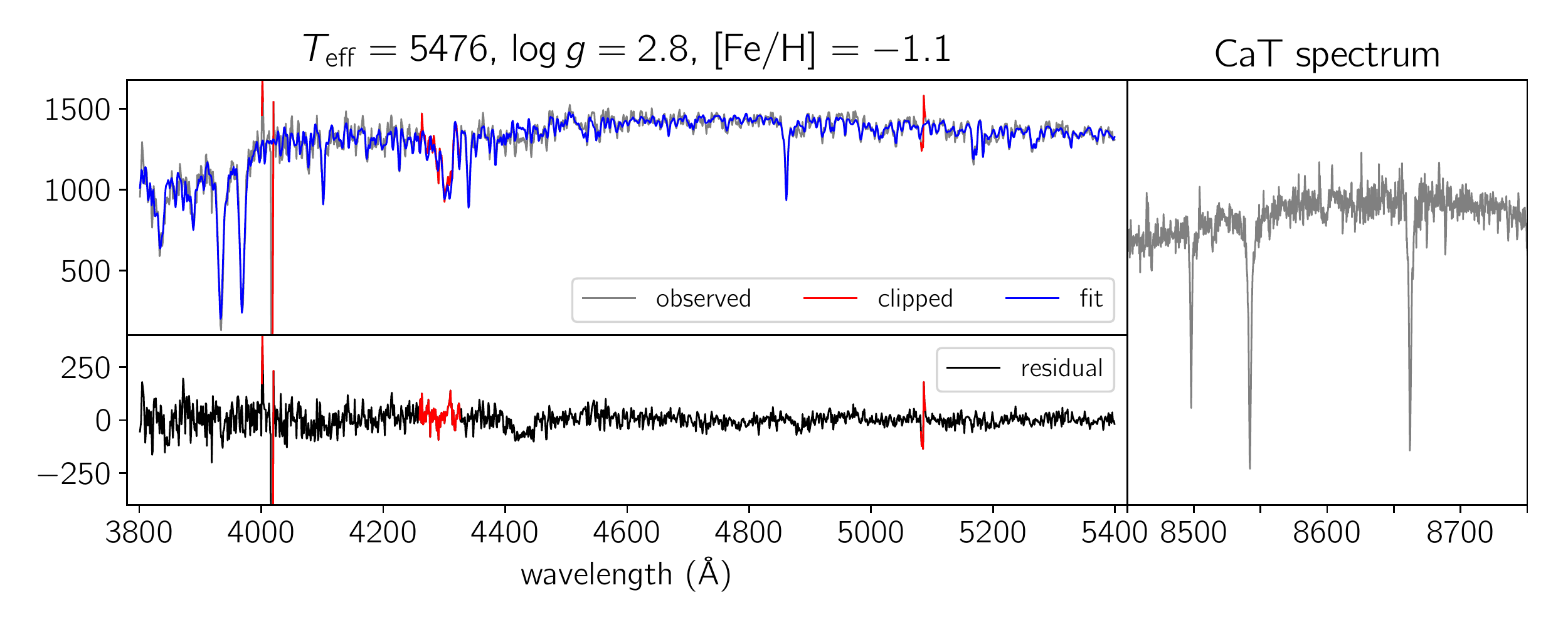}
\includegraphics[width=0.9\hsize,trim={0.5cm 0.5cm 0.0cm 0.0cm}]{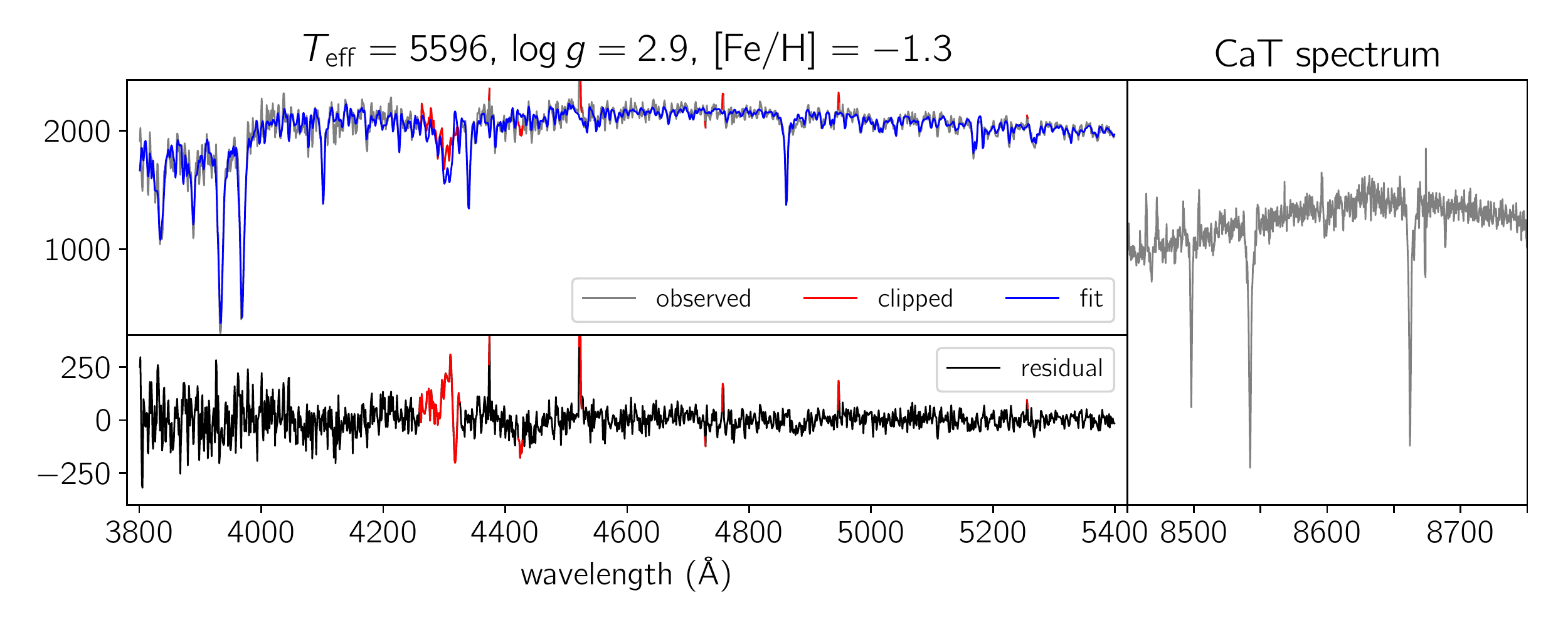}
\caption{Same as Fig.~\ref{fig:spec}, but for two HB stars, with slightly different temperatures and metallicities. }
 \label{fig:specHB}
\end{figure*}

The HB stars end up in a narrow range of metallicity around $-1.1$. Additionally the HB is not as horizontal in the \teff--\logg diagram as it should be. Both of these effects may (partly) be artefacts of the reference library. We used the MILES library \citep{sanchez,falconbarroso}, which contains only a small number of stars at \teff--\logg combinations of HB stars, that, compounding the problem, only possess these metallicities. However, preliminary tests with synthetic spectra appear to converge on metallicities similar to those from ULySS. Additionally, the ULySS fits do look reasonable, see Fig.~\ref{fig:specHB} for two examples. We therefore assume that the metallicities are good enough to separate the stars in bins of 0.5~dex as in Fig.~3 of the main paper, but we exclude them from our analysis in main paper Fig.~4, which is divided in bins of 0.25~dex. We check that when we do include them, the results do not change significantly: another indication that the metallicities are reasonable.

To separate the HB stars from normal giants, we first draw a rough box in \teff--\logg space where HB stars are located. This region also contains some normal giants, but at these temperatures they are at lower \feh than the HB stars. Therefore we choose to split the two samples within this boxed region by \feh. From the metallicity distribution in Fig.~2 of the main paper, we deduce that the \feh cut should be between \feh $ = -1.1$ and $-1.7$ (the two peaks). We assume that the two types of stars should have roughly Gaussian distributions in \feh around their mean, with some tails. The best result is obtained when using a cut at \feh $= -1.4$.

\begin{figure}
\centering
\includegraphics[width=0.67\hsize,trim={0.5cm 0.5cm 0.0cm 0.0cm}]{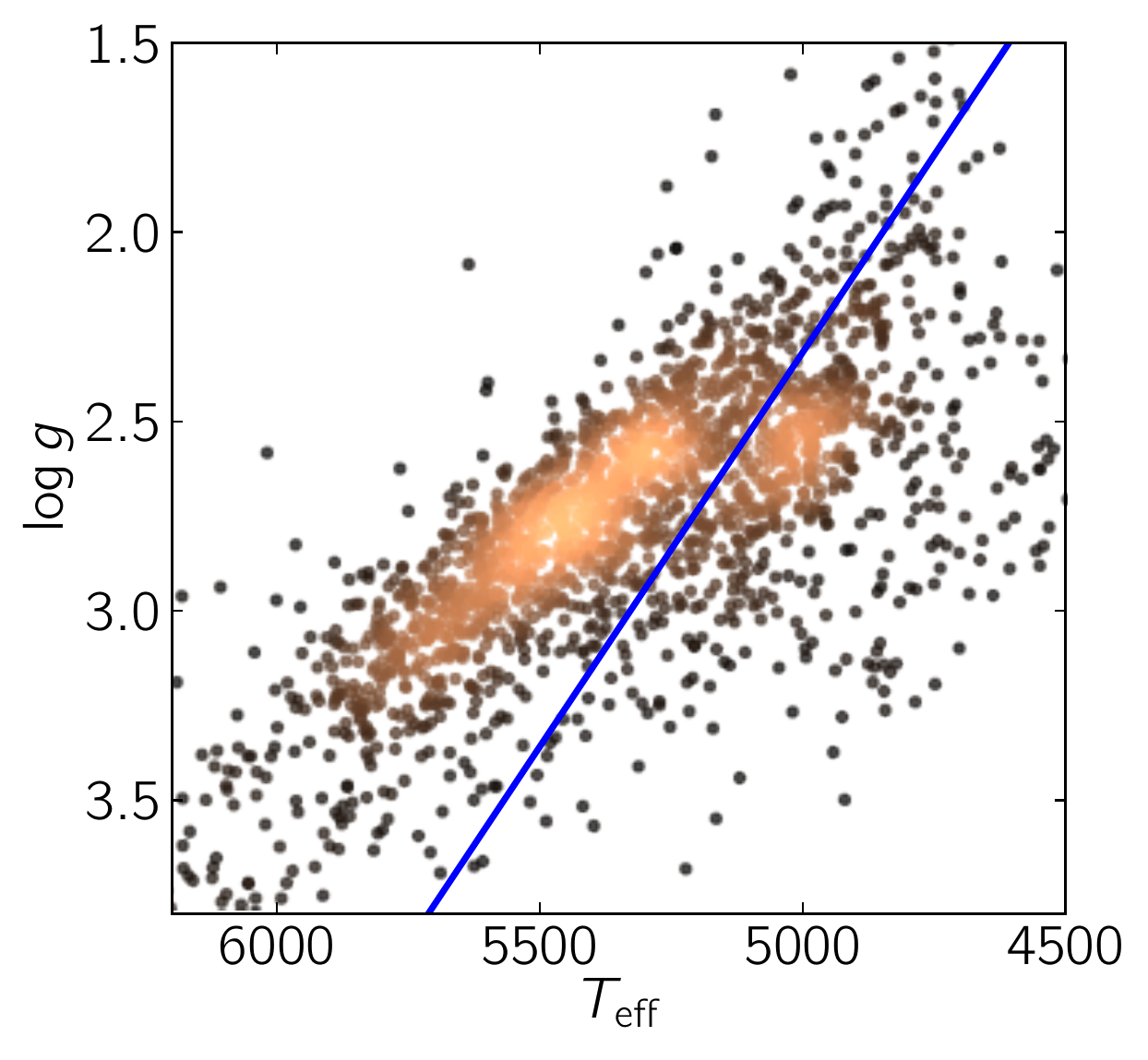}
\caption{Distribution of stars with $-1.4 < \feh < -0.5$ in the \teff--\logg diagram, colour-coded by density. We take the blue line ($\logg = \teff/480 - 8.1$) as the best separation between the normal giants and the HB stars in this metallicity range.}
 \label{fig:HBsel2}
\end{figure}

To refine the exact HB selection box further, we take the stars with $-1.4 < \feh < -0.5$ to determine what \teff--\logg combination results in the most natural split between normal and HB stars, see Fig.~\ref{fig:HBsel2}. The red line runs through the valley between the HB stars and the giants, and has been set to be oriented along the giant branch. The functional form of this line is $\logg = \teff/480 - 8.1$, which we use as the limit of our HB selection box.

\newpage
\notea{..}
\newpage
\notea{..}

\newpage

\section{Distances to our sample stars}\label{appendix_dist}

To determine whether the stars in this sample are truly ``in the bulge'' (e.g. with R$_\mathrm{GC} < 3.5$ kpc, as in \citealt{ness13b}), distances are needed. However, good distances to the stars in our sample are difficult to determine. $Gaia$ DR2 \citep{gaia18} parallaxes by themselves are not good enough to estimate precise distances to sources further away than $\sim$4--5 kpc and our uncertainties on \logg translate into large spectroscopic distance uncertainties.

Since HB stars are standard candles, we can derive distances for this subset of stars from the photometry. We use the following HB absolute magnitude relation from \citet{chen09} for the SDSS $i$-band\footnote{We convert the Pan-STARRS i-band magnitude to the SDSS photometric system using the relation from \citet{tonry12}.}:

\vspace{-0.25cm}

\begin{equation}
M_{i\mathrm{, HB}} =  \left( 0.064 \times \feh \right) +  \left(0.049 \times \mathrm{age} \right) - 0.148
\end{equation}

\noindent 
with the age given in Gyr. We assume an age of 10 Gyr and the \feh from ULySS. We convert the absolute magnitude to distance using the distance-magnitude relation: 

\vspace{-0.25cm}

\begin{equation}
d $ = $ 10^{\frac{m_i-M_i}{5} \, + \, 1}
\end{equation}

\begin{figure}
\centering
\includegraphics[width=0.9\hsize,trim={0.5cm 0.5cm 0.0cm 0.0cm}]{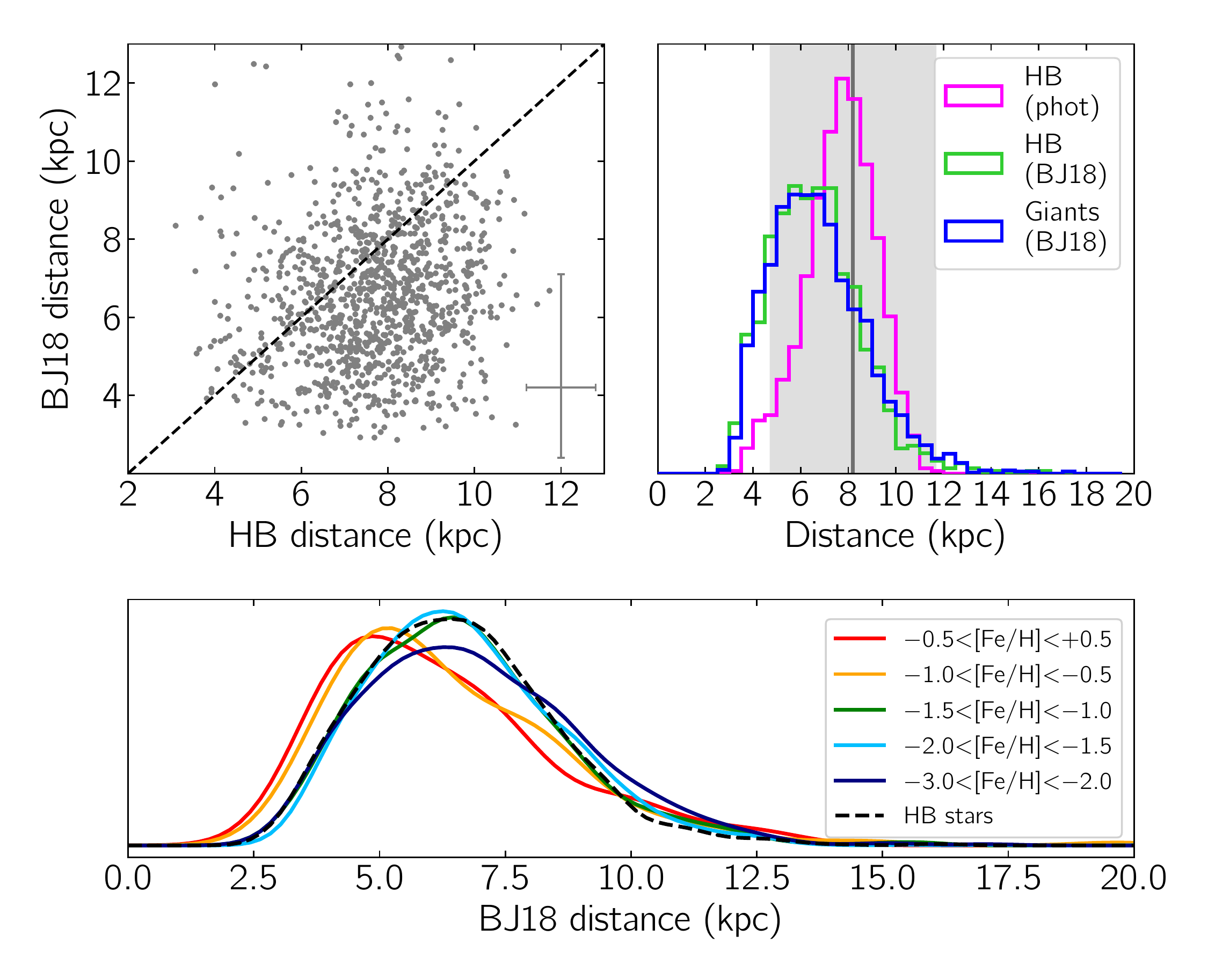}
\caption{
Top left: Comparison of the Bailer-Jones and the standard candle HB distances. A typical error bar is indicated in the lower right corner. 
Top right: Comparison of the normalised distributions of the Bailer-Jones distances for the HB stars and the giants in our sample, and the HB standard candle distances from photometry. The shaded region indicates the Galactic centre distance of 8.2 kpc \citep{gravity19} $\pm$ 3.5 kpc. 
Bottom: kernel density estimate BJ18 distance distributions for the giants in different metallicity ranges, and for the HB stars.
}
 \label{fig:dist}
\end{figure}

We derive distance uncertainties of $\sim$0.8 kpc ($\approx 10$ per cent) by quadratically adding results from a 0.2 mag uncertainty in the $i$-band (estimated to cover the uncertainty on the extinction correction and the scatter in the relation) and a 2 Gyr uncertainty in age. The dependence on metallicity is almost negligible. The resulting distance distribution is shown in magenta in the top right panel of Fig.~\ref{fig:dist}, with an average of 7.7 kpc and a Gaussian width of 1.5 kpc, they lie perfectly in the bulge region.

We compare our standard candle HB distances with distances from the $Gaia$ DR2 \citet[][hereafter BJ18]{bailerjones18} distance catalogue, which is based on parallaxes only in combination with a density prior. We remove stars which have $ruwe > 1.4$. The histograms of BJ18 distances for both the HB stars and the giants are shown in the top right panel of Fig.~\ref{fig:dist}, and the individual HB stars in the top left panel. It is clear that the BJ18 distances are generally too small and that their average uncertainties are much larger ($^{+2.9}_{-1.8}$ kpc). These distances should not be used for stars towards/in the bulge. The BJ18 distributions of the giants and HB stars look similar, indicating that these samples either occupy the same distance range, and/or are too dominated by the assumed density prior.

There is still some relative information present in the distributions. We see an effect with metallicity as shown in the bottom panel of Fig.~\ref{fig:dist}. The stars in the two most metal-rich bins have smaller BJ18 distances, consistent with them being part of the bar which is pointing towards us in this part of the sky. The most metal-poor stars with $\feh < -2.0$ may have a more pronounced tail towards larger distances. 

In a future work we will derive better individual distances, combining parallax information with spectroscopy and photometry. By themselves each of these methods have their problems and challenges (too far away for good parallax information, too large uncertainties on \logg and/or extinction for good photometric or spectroscopic distances), but together they can break the degeneracies. 

\bsp	
\label{lastpage}
\end{document}